\documentclass[
    ,final            
  ]
  {aipproc}

\layoutstyle{6x9}

\def\xmm {{\em XMM-Newton}}
\def\vlt {{\em VLT}}

\def\psr {PSR\,J2144--3933}
\def\sgr {SGR\,0418+5729}

\def\ltsima{$\; \buildrel < \over \sim \;$}
\def\lsim{\lower.5ex\hbox{\ltsima}}
\def\gtsima{$\; \buildrel > \over \sim \;$}
\def\gsim{\lower.5ex\hbox{\gtsima}}
\def\msole{~M_{\odot}}

\def\vlt{{\em VLT}}

\def\fors{{\em FORS2}}

\begin{document}

\title{The first deep X-ray and optical observations of the closest isolated radio pulsar}


\author{A. Tiengo}{
  address={INAF - Istituto di Astrofisica Spaziale e Fisica Cosmica - Milano, via E.~Bassini 15, I-20133 Milano, Italy}
}

\author{R. Mignani}{
  address={University College London, Mullard Space Science Laboratory, Holmbury St. Mary, Dorking, Surrey RH5 6NT, UK}
}

\author{A. De Luca}{
  address={INAF - Istituto di Astrofisica Spaziale e Fisica Cosmica - Milano, via E.~Bassini 15, I-20133 Milano, Italy}
  ,altaddress={IUSS - Istituto Universitario di Studi Superiori, viale Lungo Ticino Sforza 56, I-27100 Pavia, Italy} 
}

\author{P. Esposito}{
  address={INAF - Osservatorio Astronomico di Cagliari, localit{\`a} Poggio dei Pini, strada 54, I-09012 Capoterra, Italy}
}

\author{S. Mereghetti}{
  address={INAF - Istituto di Astrofisica Spaziale e Fisica Cosmica - Milano, via E.~Bassini 15, I-20133 Milano, Italy}
}

\author{A. Pellizzoni}{
  address={INAF - Osservatorio Astronomico di Cagliari, localit{\`a} Poggio dei Pini, strada 54, I-09012 Capoterra, Italy}
}

\begin{abstract}
With a distance of 170 pc, \psr\ is the closest isolated radio pulsar currently known. It is also the slowest and least energetic radio pulsar; indeed, its radio emission is difficult to account for with standard pulsar models, since its position in the $P-\dot{P}$ diagram is far beyond typical ``death lines''. Here we present the first deep X-ray and optical observations of \psr, performed in 2009 with \xmm\ and the \vlt, from which we can set one of the most robust upper limits on the surface temperature of a neutron star. We have also explored the possibility of measuring the neutron star mass from the gravitational lensing effect on a background optical source.
\end{abstract}

\maketitle


\section{Introduction}

The timing parameters of \psr\ are unique among the $\sim$2000 pulsars currently known: it is the radio pulsar with the longest spin period ($P$=8.51 s, \cite{young99}), with the lowest rotational energy loss ($\dot{E}_{\rm rot}=4\pi^2 I\dot{P}P^{-3}\simeq2.6\times10^{28}$ erg s$^{-1}$) and the farthest below the radio pulsar death line (see Figure~1, left panel), where radio emission should be inhibited (\cite{chen93}).
The interest in this source further increased when accurate distance ($172^{+20}_{-15}$ pc, corrected for the Lutz--Kelker bias \cite{verbiest10}) and proper motion ($\mu=166\pm1$  mas yr$^{-1}$) were measured with VLBI observations at the Australian Long Baseline Array \cite{deller09}. Although \psr\ is the closest isolated radio pulsar known to date, it has never been detected outside of the radio waveband.
The pulsar sky region was serendipitously observed for 5 ks with ROSAT/HRI \citep{pfeffermann87} in 1997, for 20 ks in a EUVE deep survey observation \citep{korpela98} and for 200 s by {\em GALEX} \citep{martin05} during the all-sky imaging survey. We report the results of the first deep observations of \psr\ in the X-ray and optical bands with \xmm\ and the \vlt\ (results are presented in more detail in \cite{tiengo11}).

\section{Observations and results}

\subsection{X-ray observation}

\psr\ was observed by \xmm\ for about 40 ks on 2009 October 24. After filtering out the time intervals affected by a high level of particle background, the net exposure times were 22 ks for the EPIC PN \cite{struder01} and 26 ks for the two EPIC MOS cameras \cite{turner01}. After standard event selection, X-ray images were produced in several energy bands and analyzed with different source detection algorithms, but no X-ray source was found at the position of \psr\ (see Figure~1, middle panel). The 3$\sigma$ upper limit on the source net count rate in a 10$^{\prime\prime}$ radius circle in the 0.2--10 keV energy range is 1.5$\times$10$^{-3}$ and 8.3$\times$10$^{-4}$ counts/s for the PN and the sum of the two MOS, respectively. The count rate upper limit for the PN in the 0.2-1 keV energy band, where our observation is most sensitive to the presumably soft X-ray emission of \psr, is 4.9$\times$10$^{-4}$ counts/s.

\subsection{Optical observation}

We observed \psr\  with the \vlt\ at  the ESO  Paranal observatory on  2009 August 21 with  the \fors\ blue-sensitive CCD detector at the Antu UT1 telescope. The integration times were 8850 s with the $U$ and $B$ filters, and 2950 s with the $V$ filter. As can be seen in Figure~1 (right panel), a relatively bright ($U=23.39\pm  0.05$, $B=23.82 \pm 0.03$, $V=23.76 \pm 0.05$) object is detected 1.2$^{\prime\prime}$ from the proper-motion-corrected position of \psr. Considering that the $1 \sigma$ uncertainty  of our astrometry is 0.24$^{\prime\prime}$ and that the source is slightly extended, this object is very likely unrelated to the pulsar and it is probably a background galaxy.
After correcting for atmospheric extinction and taking into account the contamination from the nearby source, the 3$\sigma$ upper limits to the emission of \psr\ are $U>25.3$, $B>26.6$, and $V>25.5$.

\begin{figure}
  \resizebox{\hsize}{!}{\includegraphics[width=.5\textwidth]{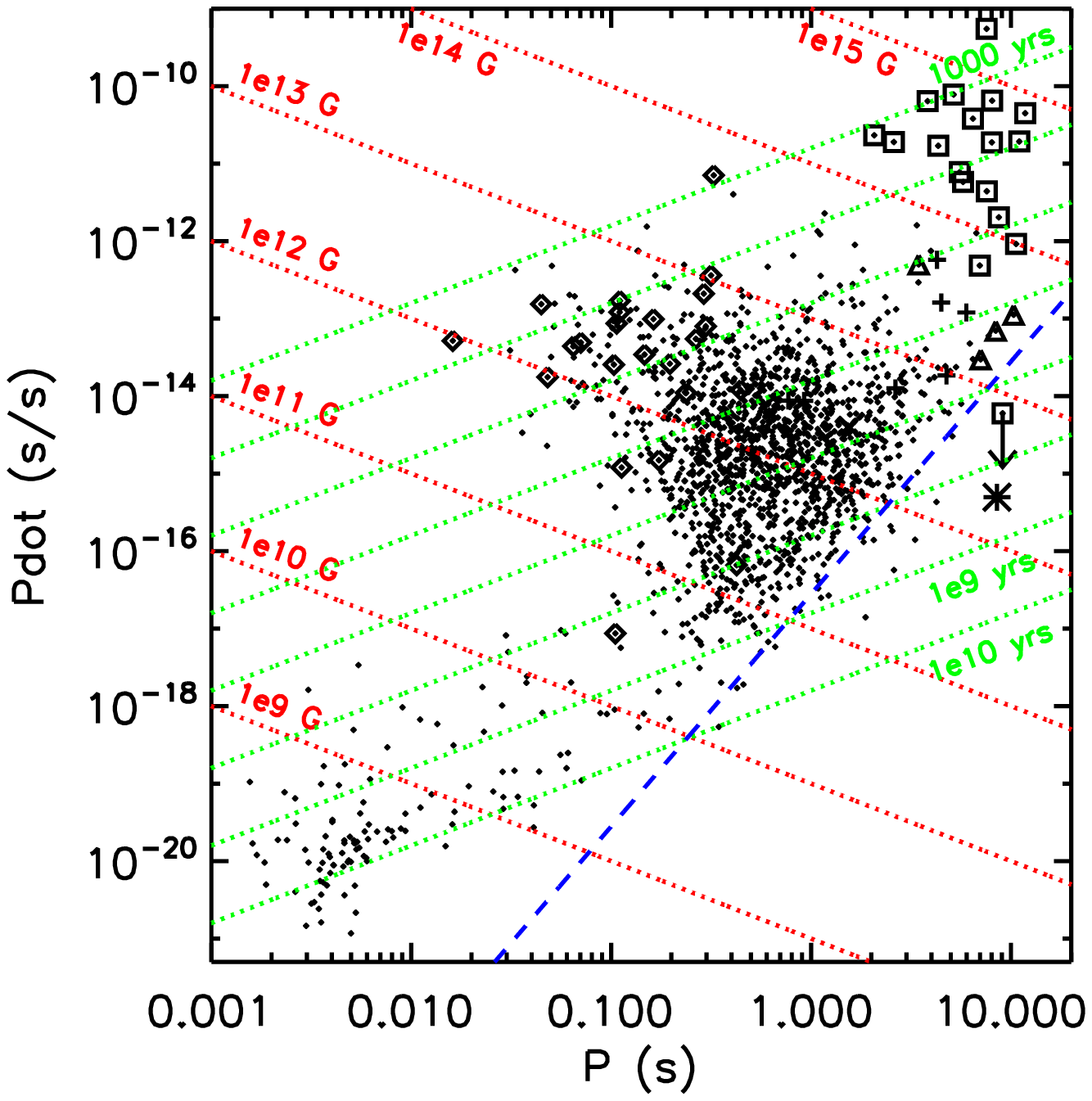}
  \includegraphics[width=.5\textwidth]{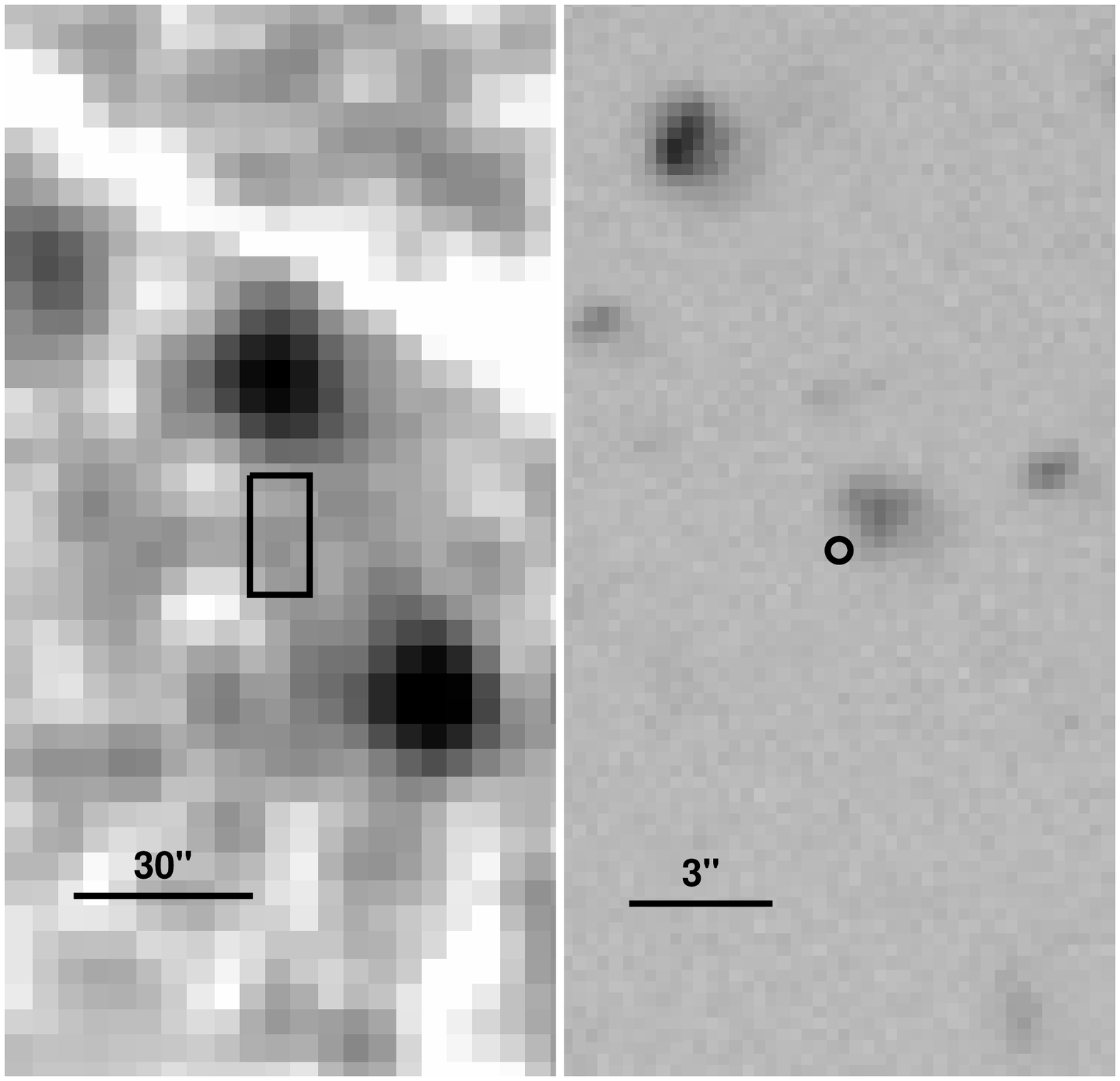}}
  \caption{{\it Left panel:} $P-\dot{P}$ diagram for radio pulsars (dots), radio-quiet $\gamma$-ray pulsars (diamonds), RRATs (crosses), XDINs (triangles), AXPs/SGRs (squares), and \psr\ (large star). Data from the ATNF Pulsar Catalogue (\cite{manchester05}, http://www.atnf.csiro.au/research/pulsar/psrcat) and from the McGill SGR/AXP Online Catalog (http://www.physics.mcgill.ca/~pulsar/magnetar/main.html). The arrow indicates the $\dot{P}$ upper limit for \sgr\ \cite{rea2010}. Dotted lines represent dipole magnetic fields from 10$^9$ to 10$^{15}$ G and characteristic ages from 10$^3$ to 10$^{10}$ years; the dashed line is a typical death line ($B/P^2=1.7\times10^{11}$ G s$^{-2}, $\cite{bhatta92}). {\it Middle panel:} \emph{XMM-Newton} PN (slightly smoothed) image of the $\sim$1.5$^{\prime}$$\times$3$^{\prime}$ field around \psr\ in the 0.2--1 keV energy range. The 10$^{\prime\prime}$$\times$20$^{\prime\prime}$ box indicates the sky region shown in the right panel. {\it Right panel:} ESO/\vlt\ B-filter image of the $\sim$10$^{\prime\prime}$$\times$20$^{\prime\prime}$ field around \psr. The pulsar proper-motion-corrected position is marked by a circle with a radius corresponding to the image astrometric uncertainty (0.24$^{\prime\prime}$).}
\end{figure}

\section{Discussion and conclusions}

Assuming a column density of $N_{\rm H}=10^{20}$ cm$^{-2}$ and an interstellar reddening $E(B-V)=0.02$, the non detection of \psr\ in our deep X-ray and optical observations corresponds to the following 3$\sigma$ upper limits on the surface temperature (measured at infinity, assuming a blackbody spectrum): $2.3\times10^5$ K for a 13 km radius neutron star, $4.4\times10^5$ K for a 500 m radius hot spot, and $1.9\times10^6$ K for a 10 m radius polar cap. The upper limits on the luminosity of the pulsar non-thermal emission (assuming a power-law spectrum with photon index $\Gamma=2$) are instead $7\times10^{27}$ erg s$^{-1}$ in the 0.5--2 keV energy band and $5.6 \times 10^{26}$ erg s$^{-1}$ in the B optical band.

Given the small and well-determined distance of \psr\ and the high quality of our observations, these limits are very robust and their values are among the lowest ever obtained for a neutron star. However, the large characteristic age ($\tau_c=P/(2\dot{P})\simeq3.4\times10^8$ years) and extremely low rotational energy loss of \psr\ make them not particularly constraining both for the study of neutron star cooling evolution and pulsar non-thermal emission processes: in fact, the surface temperature is expected to drop well below 10$^{5}$ K after $\sim$10$^7$ years and typical efficiencies of radio pulsars in converting rotational energy into the X-ray and optical radiation are much lower than our upper limits of 30\% and 2\%, respectively. Similarly, polar cap emission (for polar caps with radii $>$10 m) could have been detected by our X-ray observation only for an unprecedented polar cap emission efficiency $>$40\%.

On the other hand, the peculiar position of \psr\ in the $P-\dot{P}$ diagram (see Figure~1, left panel), in the period range of  AXPs/SGRs \cite{mereghetti08} and XDINSs \cite{haberl07}, but with a weaker dipolar magnetic field
($B=\big(\frac{3c^3I}{8\pi R^6}P\dot{P}\big)^{1/2}\simeq1.9\times10^{12}$ G),
might indicate that \psr\ was born as a magnetar, but its magnetic field has now decayed to an intensity typical of radio pulsars. In such a case, it would be significantly younger than indicated by its characteristic age and would have also been heated by the decay of its field \cite{pons09}. The recent discovery that the transient \sgr\ has a small period derivative \cite{rea2010} has shown that a pulsar with timing parameters similar to those of \psr\ (see the left panel of Figure~1) can be a bright X-ray source.

The background galaxy that worsened our sensitivity to detect \psr\ in the \vlt\ images, might potentially be a tool to measure the neutron star mass through gravitational lensing (see, e.g., \cite{paczynski96hst}). However, the displacement of a background source $\Delta \varphi=1.2^{\prime\prime}$ away from a $M=1.4 \msole$ neutron star at a distance $D=170$ pc is expected to be only $\delta \varphi =\frac{4GM}{c^2D\Delta \varphi}\simeq0.06$ mas, which is beyond the capability of existing optical instruments. The situation is not going to improve in the next years, since the pulsar proper motion is directed away from the nearby source.
A second gravitational lensing effect that would be in principle observable in this case, is the formation of a secondary ghost image of the galaxy close to the position of the pulsar, with a flux smaller than the one of the original source by a factor $A_-=0.5-\frac{u^2+2}{2u(u^2+4)^{1/2}}$,
 where $u=\Delta \varphi\big(\frac{4GM}{c^2 D}\big)^{-1/2}$. For a  canonical neutron star mass, the $B=23.8$ magnitude background galaxy would have a secondary image of magnitude $B = 45$, while only a lens of at least
2$\times$10$^4 \msole$ would produce a detectable image in our \vlt\ data!
Not a particularly constraining upper limit for a neutron star mass, but the first one reported to date with this method...
%
%
%

A detectable lensing effect, that would lead to a measure of the neutron star mass, would instead be possible in case deeper observations of this field will be able to find dimmer sources along the sky trajectory of \psr.
This pulsar is an ideal target for this challenging measurement thanks to its small distance, its very well determined position and proper motion and its vanishing flux outside the radio band.

\begin{theacknowledgments}
We acknowledge the partial support from ASI (ASI/INAF contract I/088/06/0). PE acknowledges financial support from the Autonomous Region of Sardinia through a research grant under the program PO Sardegna FSE 2007--2013, L.R. 7/2007.
\end{theacknowledgments}



\bibliographystyle{aipproc}   




\end{document}